\documentclass[conference, 10pt, letterpaper]{IEEEtran}
\IEEEoverridecommandlockouts
\usepackage{graphicx}
\usepackage{cite}
\usepackage{amsmath,amssymb,amsfonts}
\usepackage{textcomp}
\usepackage{xcolor}
\usepackage{array}
\usepackage{arydshln}
\usepackage{url}
\usepackage[hidelinks, bookmarks=false]{hyperref}
\usepackage{algorithm}
\usepackage[noend]{algpseudocode}
\makeatletter
\algrenewcommand\ALG@beginalgorithmic{\footnotesize}
\algnewcommand\And{\textbf{and}}
\makeatother
\usepackage{multirow}
\usepackage[flushleft]{threeparttable}
\usepackage[caption=false]{subfig}
\newcolumntype{P}[1]{>{\centering\arraybackslash}p{#1}}
\newcolumntype{M}[1]{>{\centering\arraybackslash}m{#1}}
\usepackage{balance}
\usepackage[draft]{todonotes}
    
\begin{document}

\title{Transfer Learning for Piano Sustain-Pedal Detection \\
\thanks{This work is supported by Centre for Doctoral Training in Media and Arts Technology (EPSRC and AHRC Grant EP/L01632X/1), the EPSRC Grant EP/L019981/1 ``\textit{Fusing Audio and Semantic Technologies for Intelligent Music Production and Consumption (FAST-IMPACt)}'' and the European Commission H2020 research and innovation grant AudioCommons (688382). Beici Liang is funded by the China Scholarship Council (CSC).}
}

\author{\IEEEauthorblockN{Beici Liang, Gy\"{o}rgy Fazekas and Mark Sandler}
\IEEEauthorblockA{Centre for Digital Music, Queen Mary University of London\\
London, United Kingdom\\
Email: \{beici.liang,g.fazekas,mark.sandler\}@qmul.ac.uk}}

\maketitle

% IJCNN: Length of abstract has to be between 200 and 1750 characters.
\begin{abstract}
Detecting piano pedalling techniques in polyphonic music remains a challenging task in music information retrieval. While other piano-related tasks, such as pitch estimation and onset detection, have seen improvement through applying deep learning methods, little work has been done to develop deep learning models to detect playing techniques. In this paper, we propose a transfer learning approach for the detection of sustain-pedal techniques, which are commonly used by pianists to enrich the sound. In the source task, a convolutional neural network (CNN) is trained for learning spectral and temporal contexts when the sustain pedal is pressed using a large dataset generated by a physical modelling virtual instrument. The CNN is designed and experimented through exploiting the knowledge of piano acoustics and physics. This can achieve an accuracy score of 0.98 in the validation results. In the target task, the knowledge learned from the synthesised data can be transferred to detect the sustain pedal in acoustic piano recordings. A concatenated feature vector using the activations of the trained convolutional layers is extracted from the recordings and classified into frame-wise pedal press or release. We demonstrate the effectiveness of our method in acoustic piano recordings of Chopin's music. From the cross-validation results, the proposed transfer learning method achieves an average F-measure of 0.89 and an overall performance of 0.84 obtained using the micro-averaged F-measure. These results outperform applying the pre-trained CNN model directly or the model with a fine-tuned last layer. 
\end{abstract}

% IJCNN: If you are using the Latex template, do not include keywords in your paper.
% \begin{IEEEkeywords}
% playing technique detection, convolutional neural networks, transfer learning.
% \end{IEEEkeywords}

% \todo[inline]{REWIEWERS' MAIN COMMENTS: \\
% 1. SVM is not motivated enough. \\
% 2. No data regarding the performance of the CNN if we conduct a short incremental training of the fully connected layers. \\
% 3. I think the authors are wrong in giving so much weight to an improvement on the 4th decimal and should just say that the convnet-multi is taken because numerically it is slightly better, without so much emphasis on the weak performance boost. \\
% 4. More clarification for the evaluation setup.
% }

% ==============================
%  Introduction
% ==============================
\section{Introduction}
% Mastering the use of pedals strongly relies on listening to nuances in piano sound. To develop critical listening, piano students need instructions with respect to when the pedal should be pressed and for what duration. Most of the advice in piano lessons refer to key touch however, while relatively little attention is usually dedicated to the use of pedals \cite{schnabel1954modern}. This is despite the fact that acquiring pedalling techniques merely by experimentation can be fairly time consuming and may not lead to full exploitation of an expressive vocabulary granted by the instrument. To facilitate the learning process, we pose a research question: ``Can a computer point out pedalling techniques when a piano recording from a virtuoso performance is given?'' 

Learning to use the piano pedals strongly relies on listening to nuances in the sound. Instructions with respect to when the pedal should be pressed and for what duration are required to develop critical listening. To facilitate the learning process, we pose a research question: ``Can a computer point out pedalling techniques when a piano recording from a virtuoso performance is given?'' Pedalling techniques change very specific acoustic features, which can be observed from their spectral and temporal characteristics on isolated notes. However, their effects are typically obscured by the variations in pitch, dynamics and other elements in polyphonic music. Therefore, automatic detection of pedalling techniques using hand-crafted features is a challenging problem. Given enough labelled data, deep learning models have shown the ability of learning hierarchical features. If these features are able to represent acoustic characteristics corresponding to pedalling techniques, the model can serve as a detector.

\begin{figure}[t]
\centering
\includegraphics[width=.85\columnwidth]{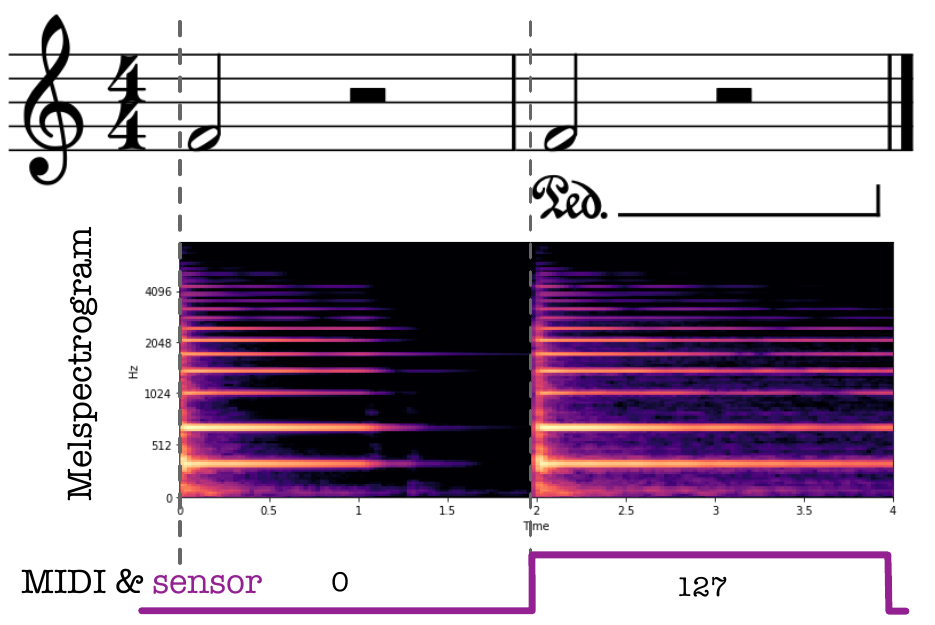}
\caption{Different representations of the same note played without (first note) or with (second note) the sustain pedal, including music score, melspectrogram and messages from MIDI or sensor data.}
\label{fig:intro}
\end{figure}

In this paper, we focus on detecting the technique of the sustain pedal, which is the most frequently used one among the three standard piano pedals. All dampers are lifted off the strings when the sustain pedal is pressed. This mechanism helps to sustain the current sounding notes and allows strings associated to other notes to vibrate due to coupling via the bridge. A phenomenon known as \textit{sympathetic resonance} \cite{morfey2001dictionary} is thereby enhanced and embraced by pianists to create a ``dreamy'' sound effect. We can observe how the phenomenon reflects on the melspectrogram in Figure \ref{fig:intro}, where note \textit{F4} is played without (first) and with (second) the sustain pedal in two bars respectively. Note that the symbol under the second bar of the music score in Figure \ref{fig:intro} can be used to indicate the sustain-pedal techniques. Yet, even if pedal notations are provided, pedalling in the same piano passage can be executed in many different ways. Playing techniques are typically adjusted to the performer's sense of tempo, dynamics, as well as the location where the performance takes place \cite{rosenblum1993pedaling}.

% It is noted that the symbol under the second bar of the music score in Figure \ref{fig:intro} can be used to indicate the sustain-pedal techniques. Composers like Chopin take particular care of pedal notations, allowing performers to receive sufficient indications for their interpretations. Yet, pedal notations can be scarce in the scores of composers such as Debussy, despite the importance of pedalling in the performance of their works. Even if pedal notations are provided, pedalling in the same piano passage can be executed in many different ways. Playing techniques are typically adjusted to the performer's sense of tempo, dynamics, as well as the location where the performance takes place \cite{rosenblum1993pedaling}.

Given that detecting pedalling nuances from the audio signal alone is a rather challenging task \cite{goebl2008sense}, several measurement systems have been developed to capture the pedal movement. For instance, the Yamaha Disklavier piano can encode this movement into MIDI messages (0-127) along with note events. A dedicated system proposed in \cite{liang2018measurement} enables synchronously recording the pedalling gestures and the piano sound. This can be deployed on common acoustic pianos, and it is used to provide the ground truth dataset introduced in Section \ref{sec:dataset}. 

\begin{figure}[t]
\centering
\includegraphics[width=.85\columnwidth]{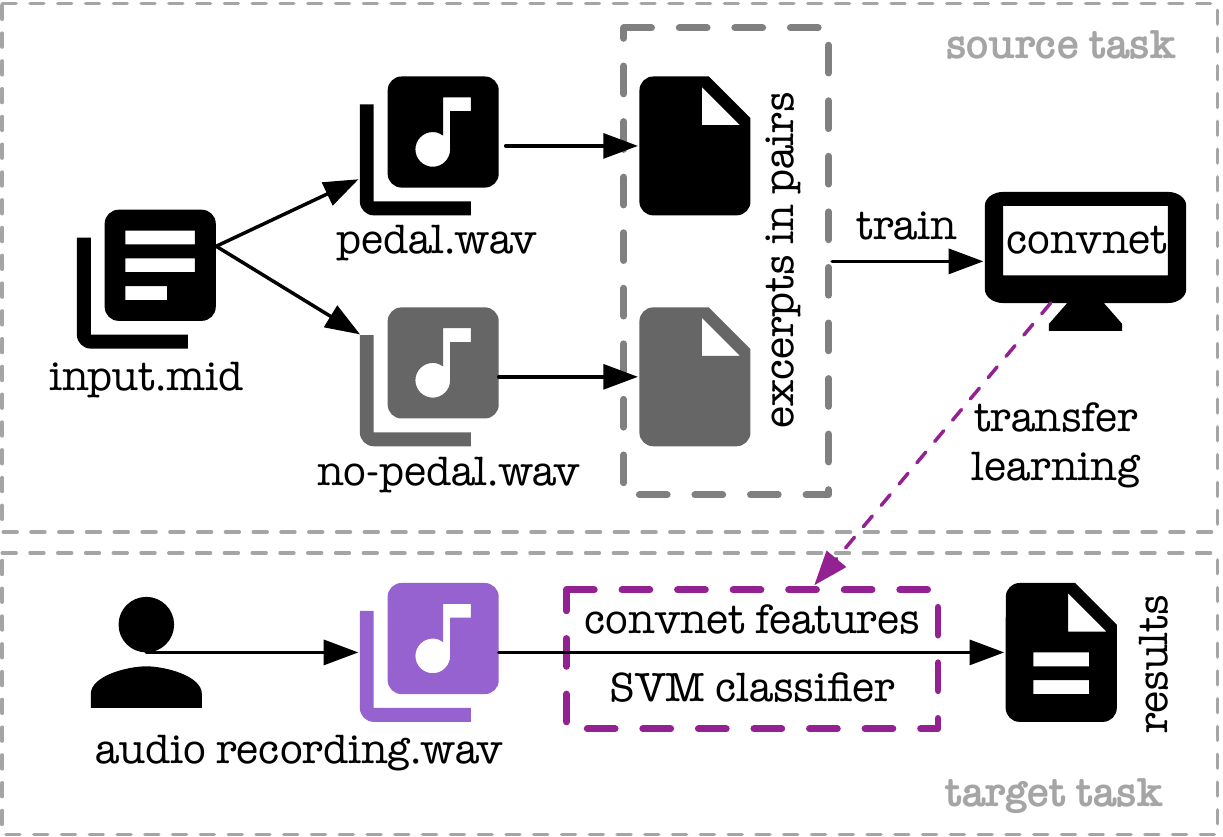}
\caption{Framework of the proposed method.}
\label{fig:framework}
\end{figure}

Detection of pedalling techniques from audio recordings is necessary in the cases where installing sensors on the piano is not practical. We approach the sustain-pedal detection from the audio domain using transfer learning \cite{goodfellow2016deep} as illustrated in Figure \ref{fig:framework}. Transfer learning exploits the knowledge gained during training on a source task and applies this to a target task \cite{pan2010survey}. This is crucial for our case, where the target-task data is obtained from recordings of a different piano, therefore it is difficult to learn a ``good'' representation due to mechanical and acoustical deviations. In our source task, a convolutional neural network (denoted by \texttt{convnet} hereafter) is trained for distinguishing synthesised music excerpts with or without the sustain-pedal effect. The \texttt{convnet} is then used as a feature extractor, aiming to transfer the sustain-pedal effect learned from the source task to the target task. Support vector machines (SVMs) \cite{suykens1999least} are trained using the frame-wise \texttt{convnet} features from the acoustic piano recordings to finalise the feature representation transfer as the target task. SVMs can be used as a classifier to localise which frames are played with the sustain pedal. The performance is expected to improve significantly with the new feature representation. To sum up, the main contributions of this paper are:
\begin{enumerate}
  \item A novel strategy of model design, which incorporates knowledge of piano acoustics and physics, enabling the \texttt{convnet} to become more effective in representing the sustain-pedal effect.
  %with more representative power dedicated to the sustain-pedal effect.
  \item A transfer learning method that allows the \texttt{convnet} trained from the source task to be adapted to the target task, where the recording instruments and room acoustics are different. This also allows effective learning with a smaller dataset.
  \item Finally, we conduct visual analysis on the convolutional layers of the \texttt{convnet} to promote model designs with fewer trainable parameters, while maintaining their discriminating power.
\end{enumerate}

The rest of this paper is organised as follows. We first introduce related works in Section \ref{sec:relatedwork}. The process of database construction is described in Section \ref{sec:dataset}. The methods of sustain-pedal detection including \texttt{convnet} design and transfer learning are discussed in Section \ref{sec:method}. Experiments and results are presented in Section \ref{sec:experiment}. We finally conclude our work in Section \ref{sec:conclusion}.

% ==============================
%  Related Work
% ==============================
\section{Related Work}
\label{sec:relatedwork}
Past research in music information retrieval (MIR) abound in recognition of musical instruments, but automatic detection of instrumental playing techniques (IPT) remains underdeveloped \cite{lostanlen2018extended}. IPT creates a variety of spectral and temporal variations of the sounds in different instruments. Recent research has attempted to transcribe IPT on drum \cite{wu2018review}, erhu \cite{yang2017filter}, guitar \cite{su2014sparse,chen2015electric} and violin \cite{li2015analysis,perez2015indirect}. Hand-crafted features are commonly designed based on instrument acoustics to capture the salient variations induced by IPT. The sustain-pedal technique leads to rather subtle variations, therefore most studies managed to detect the technique based on isolated notes only \cite{lehtonen2007analysis,badeau2008piano,liang2017detection}. This challenge is further intensified in polyphonic music where clean features extracted from isolated notes cannot be easily obtained. In our prior work \cite{liang2018legato}, the first research aiming to extract pedalling technique in polyphonic piano music, we proposed a method for detecting pedal onset times using a measure of sympathetic resonance. Yet, this method assumes the availability of modelling the specific acoustic piano which is also used in evaluation. Moreover, it is prone to errors due to its reliance on note transcription.

Convolutional Neural Networks (CNNs) have been used to boost the performance in MIR tasks, with the ability to efficiently model temporal features \cite{pons2017designing} and timbre representations \cite{pons2017timbre}. We choose CNNs to facilitate learning time-frequency contexts related to the sustain pedal, using synthesised excerpts in pairs (\textit{pedal} versus \textit{no-pedal} versions). Using this method, contexts that are invariant to large pitch and dynamics changes can be learned.

To apply a \texttt{convnet} trained from the synthesised data into the context of real recordings, a transfer learning approach can be used. It has been gaining more attentions in MIR for alleviating the data sparsity problem and its ability to be used for different tasks. For example, Choi et al. \cite{choi2017transfer} obtained features from CNNs, which were trained for music tagging in the source task. These features outperformed MFCC features in the target tasks, such as genre and vocal/non-vocal classification. We believe such strategy is suited to the challenges in detecting the sustain pedal from polyphonic piano music recorded in different acoustic and recording conditions.

In our case, training a \texttt{convnet} with the synthesised data is considered as the source task. Then in the target task, we can use the learnt representations from the trained \texttt{convnet} as features, which are extracted from every frame of a real piano recording, to train a dedicated classifier adapted to the actual acoustics of the piano and the performance venue used in the recording. This transfer learning approach is expected to better identify frames played with the sustain pedal. For the dedicated classifier in the target task, we opt for SVM instead of multi-layer perceptron because SVM can greatly reduce the training time and yield better generalisation in classification tasks~\cite{osowski2004mlp}. In Section~\ref{sec:tt}, compared with fine-tuning the last layer of the pre-trained \texttt{convnet}, transfer learning with SVM trained using the activations of multiple layers also achieves better performance.

% ==============================
%  Dataset
% ==============================
\section{Dataset}
\label{sec:dataset}
For the source task, \textit{pedal} and \textit{no-pedal} versions of music excerpts are required to train a \texttt{convnet}, which is able to highlight the spectral or temporal characteristics that change with the sustain pedal instead of note events. For this reason, 1392 MIDI files publicly available from the Minnesota International Piano-e-Competition website\footnote{\url{http://www.piano-e-competition.com}} were downloaded. They were recorded using a Yamaha Disklavier piano from the performance of skilled competitors. To render these MIDI files into high quality audio, the Pianoteq 6 PRO\footnote{\url{https://www.pianoteq.com/pianoteq6}} software was used. This physically modelled virtual instrument approved by Steinway \& Sons can export audio using models of different instruments and recording conditions. We employed the Steinway Model D grand piano instrument and the close-miking recording mode. Audio with or without sustain-pedal effect was then generated with a sampling rate of 44.1 kHz and a resolution of 24 bits. These were rendered while preserving or removing the sustain-pedal message in the MIDI data. For each \textit{pedal}-version audio, we can obtain the temporal regions when the sustain pedal is \textit{on} or \textit{off} by thresholding the MIDI message at 64 given its range of [0,127]. A pedalled segment is determined to start at a pedal onset (where the pedal state changes from \textit{off} to \textit{on}) and finish when the state returns to \textit{off}. We can clip all the pedalled segments to form the \textit{pedal} excerpts. The start and end times of the pedalled segments were also used to obtain \textit{no-pedal} excerpts from the corresponding \textit{no-pedal}-version of the audio.

These music excerpts were derived from pieces by 84 different composers from Baroque to the Modern period. Their durations distribute between 0.3 and 2.3 seconds. To prepare fixed-length data for training, excerpts that are shorter or longer than 2 seconds were repeated or trimmed to create a 2-second excerpt. Considering the large size of our dataset, we randomly took a thousand samples from the excerpts of each composer. In total, 62424 excerpts form a smaller dataset\footnote{There are less than a thousand excerpts for some of the composers. The excerpts were sampled in pairs such that the ratio of \textit{pedal} and \textit{no-pedal} excerpts is 1:1.}. This also helps to compare \texttt{convnet} of different architectures in a more efficient way, since the training time can be significantly reduced.

For the target task, the dataset consists of ten well known passages of Chopin's piano music. A pianist was asked to perform the passages using a Yamaha baby grand piano situated in the MAT studios at Queen Mary University of London. The audio were recorded at 44.1 kHz and 24 bits using the spaced-pair stereo microphone technique with a pair of Earthworks QTC40 omnidirectional condenser microphones positioned about 50 cm above the strings. The positions were kept constant during the recording. Meanwhile, movement of the sustain pedal was recorded along with the audio with the help of the measurement system proposed in \cite{liang2018measurement}. The audio data were annotated with frame-wise \textit{on} or \textit{off} labels as the ground truth, representing whether the sustain pedal was pressed or released in each audio frame. The occurrence counts of the labels in each passage are presented in Table \ref{tab:tt}. It can be observed that the sustain pedal was frequently used for the interpretation of Chopin's music.

% ==============================
%  Method
% ==============================
\section{Method}
\label{sec:method}

\subsection{CNN for binary classification}
\begin{figure}[t]
\centering
\includegraphics[width=.9\columnwidth]{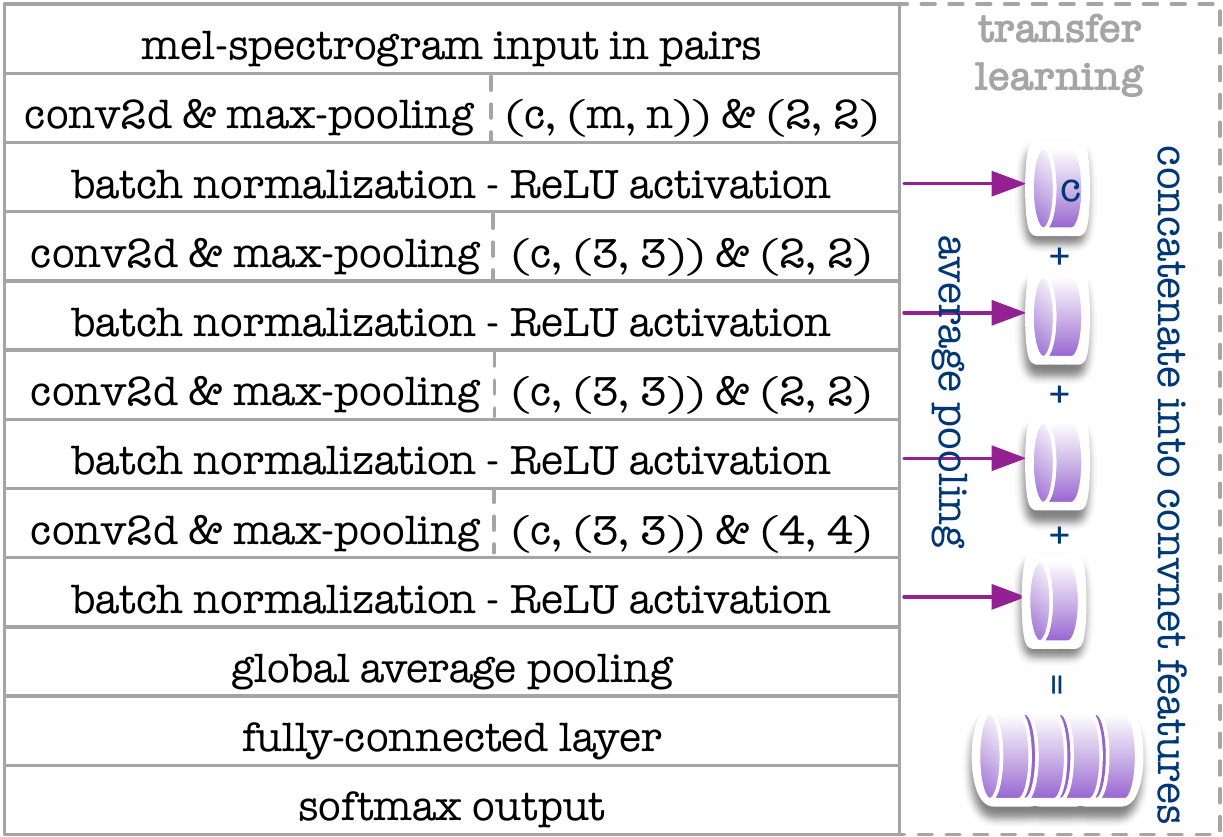}
\caption{Details of the \texttt{convnet} architecture and a schematic of feature extraction procedures during transfer learning.}
\label{fig:method}
\end{figure}
Given our large training data consisting of excerpts arranged in \textit{pedal}/\textit{no-pedal} pairs, binary classification was chosen as a source task. This enabled the \texttt{convnet} to focus on variations in the nuances on sound played with/without the sustain pedal, while invariant to other musical elements such as pitch and loudness. Considering that the use of the sustain pedal can have effects on every piano string, this could lead to changes that affect the entire spectrum, i.e., take place at a global level. Therefore representations that reveal finer details, such as short-time Fourier transform (STFT), may become inefficient for training. The melspectrogram is a 2D representation that approximates human auditory perception through aggregating STFT bins along the frequency axis. This computationally efficient input has been shown to be successful in MIR tasks such as music tagging \cite{choi2016automatic}. For the above reasons, we consider melspectrogram an adequate input representation.

Inspired by \textit{Vggnet} \cite{simonyan2014very} which has been found to be effective in music classification \cite{choi2016convolutional}, our \texttt{convnet} model uses a similar architecture with fewer trainable parameters to learn the differences in time-frequency patterns in \textit{pedal} versus \textit{no-pedal} cases. The model consists of a series of convolutional and max-pooling layers, which are followed by one fully-connected layer with two softmax outputs. The architecture we propose to start with, and the related hyperparameters are summarised in Figure \ref{fig:method}, where ($c$, ($m$, $n$)) correspond to \textit{(channel, (kernel lengths in frequency, time))} specifying the convolutional layers. Pooling layer is specified by \textit{(pooling length in frequency, time)}. 

It was noted in \cite{pons2017timbre} that designing filter shapes within the first layer can be motivated by domain knowledge in order to efficiently learn musically relevant time-frequency contexts with spectrogram-based CNNs.  To decide ($m$, $n$) of the first layer yielding the best representational power, we selected their values motivated by piano acoustics and physics, which can substantially change the sustain-pedal effect. Performance of \texttt{convnet} with different filter shapes within the first layer were evaluated using the validation set as discussed in Section \ref{sec:st}. Apart from the common small-square filter shape, the shapes we experimented with are either wider rectangles in the time domain to model short time-scale patterns, or in the frequency domain to fit spectral contexts.

In every convolutional layer, batch normalisation was used to accelerate convergence. The output was then passed through a Rectified Linear Unit (ReLU) \cite{nair2010rectified}, followed by a max-pooling layer to prevent the network from over-fitting and to be invariant to small shifts in time and frequency. To further minimise over-fitting, global average pooling was used before the final fully-connected layer. The final layer used softmax activation in order to map the output to the range [0,1], which can be interpreted as a likelihood score of the presence of the sustain pedal in the input. We trained \texttt{convnet} with the Adam optimiser \cite{kingma2014adam} to minimise binary cross entropy. 

There are possibilities that simpler model architecture, i.e., with fewer channels or convolutional layers, would be sufficient for our binary classification task using reduced parameters. We explored the effect of number of channels and layers in Section \ref{sec:st}. The best performing \texttt{convnet} model was selected to ensure the features extracted from it can accurately represent the acoustic effects when the sustain pedal is used.

\subsection{Transfer Learning}
\label{sec:tf}
When the detection aims at real piano recordings, relying on the output from the trained \texttt{convnet} may be inadequate. This is because our \texttt{convnet} was trained solely on synthesised excerpts in pairs. Only the hierarchical features representing acoustic characteristics when the sustain pedal of a virtual piano is played in the specified recording environment can be learned. It has been well understood that piano sounds can be varied by brands, and also affected by room acoustics and recording conditions. Such differences could bring more variations to the sustain-pedal effect. These serve as motivations for the proposed transfer learning, which could extract the hierarchical knowledge (specialised features) from the \texttt{convnet}. The knowledge is then used as features to train a dedicated classifier for detecting the sustain pedal of a specific piano in real scenarios.

The activations of each intermediate layers were sub-sampled using average pooling and then concatenated into the final \texttt{convnet} features as demonstrated in Figure \ref{fig:method}. Here average pooling can summarise the global statistics and reduce the size of feature maps to a vector of length associated to the value of $c$. In the end, a $c \times 4$ dimensional feature vector was generated since there are 4 convolutional layers in the \texttt{convnet}. For the brevity of this paper, the effects of using various strategies for layer-wise feature combination are not discussed.

To identify which audio frames were played with the sustain pedal, we can use SVMs to classify the frame-wise \texttt{convnet} features into pedal \textit{on} or \textit{off} states. SVMs were chosen first because we assume the features extracted from the carefully-trained model in the source task should be representative and separable. Second, the SVM algorithm was originally devised for classification problems, involving finding the maximum margin hyperplane that separates two classes of data and has been shown ideal for such a task \cite{duda2000pattern}. This allows us to focus on the quality of learnt features. SVMs were trained using a supervised learning method in the target task, where the detection was done on acoustic piano recordings. 

As shown in Section \ref{sec:tt}, the proposed transfer learning method overall outperformed the case of using the pre-trained \texttt{convnet} output directly. It also provided better performance than using the pre-trained \texttt{convnet} with a fine-tuned last layer, which is a common approach to transfer learning.

% ==============================
%  Experiment
% ==============================
\section{Experiment}
\label{sec:experiment}

\begin{table}[t]
\centering
\caption{Performance of different \texttt{convnet} models.}
\label{tab:convnet}
\begin{tabular}{|p{0.33\columnwidth}|M{0.13\columnwidth}|M{0.13\columnwidth}|M{0.13\columnwidth}| }
\hline
\textbf{Model} & ($\boldsymbol{m}$, $\boldsymbol{n}$) & \textbf{Accuracy} & \textbf{AUC}\\
\hline
\texttt{convnet-baseline} & (3, 3) & 0.9755 & \textbf{0.9963}\\
\hline
\multirow{3}{*}{\texttt{convnet-frequency}} & (9, 3) & 0.9630 & 0.9905\\\cdashline{2-4}
& (20, 3) & 0.9751 & 0.9956\\\cdashline{2-4}
& (45, 3) & 0.9747 & \textbf{0.9968}\\\hline
\multirow{3}{*}{\texttt{convnet-time}} & (3, 10) & 0.9815 & \textbf{0.9973}\\\cdashline{2-4}
& (3, 20) & 0.9787 & 0.9972\\\cdashline{2-4}
& (3, 30) & 0.9816 & 0.9971\\\hline
\end{tabular}
\end{table}

\begin{table}[t]
\centering
\caption{Performance of different models based on \texttt{convnet-multi}.}
\label{tab:convnetmulti}
\begin{threeparttable}
\begin{tabular}{|M{0.26\columnwidth}|M{0.075\columnwidth} : M{0.075\columnwidth} |M{0.13\columnwidth}|M{0.13\columnwidth}| }
\hline
\textbf{\texttt{convnet-multi}} & $\boldsymbol{c}$ & $\boldsymbol{l}$ & \textbf{Accuracy} & \textbf{AUC}\\
\hline
\multirow{8}{*}{\parbox{0.25\columnwidth}{Models with Reduced Parameters}} & 3 & 2 & 0.8781 & 0.9486\\\cdashline{2-5}
& 12 & 2 & 0.9389 & 0.9804\\\cdashline{2-5}
& 21 & 2 & 0.9552 & 0.9890\\\cdashline{2-5}
& 3 & 3 & 0.9436 & 0.9849\\\cdashline{2-5}
& 12 & 3 & 0.9708 & 0.9948\\\cdashline{2-5}
& 21 & 3 & 0.9741 & 0.9960\\\cdashline{2-5}
& 3 & 4 & 0.9513 & 0.9870\\\cdashline{2-5}
& 12 & 4 & 0.9762 & 0.9964\\\hline
Original Model & 21 & 4 & \textbf{0.9837} & \textbf{0.9983}\\
\hline
\end{tabular}
\begin{tablenotes}
      \small
      \item \textit{Note:} $l$ denotes the number of convolutional layers. 
\end{tablenotes}
\end{threeparttable}
\end{table}

\begin{table*}[t]
\centering
\caption{Performance of the two methods in the target task.}
\label{tab:tt}
\begin{tabular}{|p{0.22\columnwidth}|M{0.13\columnwidth}:M{0.13\columnwidth}|M{0.13\columnwidth}:M{0.13\columnwidth}:M{0.13\columnwidth}|M{0.13\columnwidth}:M{0.13\columnwidth}:M{0.13\columnwidth}|}
\hline
\multirow{2}{*}{\textbf{Music Passages}} & \multicolumn{2}{c|}{\textbf{Occurrence Counts}} & \multicolumn{3}{c|}{\textbf{Retrain Last Layer Only}} & \multicolumn{3}{c|}{\textbf{Transfer Learning with SVM}}\\
\cdashline{2-9}
 & \textit{on} & \textit{off} & $P_1$ & $R_1$ & $F_1$ & $P_1$ & $R_1$ & $F_1$\\\hline
 Op.10 No.3 & 849 & 268 & 0.7615 & 0.9965 & 0.8633 & 0.8457 & 0.9941 & \textbf{0.9139} \\\hdashline
 Op.23 No.1 & 722 & 355 & 0.6670 & 0.8573 & 0.7503 & 0.8643 & 0.9349 & \textbf{0.8982} \\\hdashline
 Op.28 No.4 & 995 & 322 & 0.7569 & 0.9698 & 0.8502 & 0.8148 & 0.9859 & \textbf{0.8922} \\\hdashline
 Op.28 No.6 & 788 & 289 & 0.7357 & 0.9607 & 0.8332 & 0.8178 & 0.9569 & \textbf{0.8819} \\\hdashline
 Op.28 No.7 & 291 & 66 & 0.8217 & 0.8866 & 0.8529 & 0.8971 & 0.8385 & \textbf{0.8668} \\\hdashline
 Op.28 No.15 & 611 & 306 & 0.6659 & 0.9329 & 0.7771 & 0.8412 & 0.9624 & \textbf{0.8977} \\\hdashline
 Op.28 No.20 & 783 & 274 & 0.7405 & 0.9949 & 0.8490 & 0.7849 & 0.9974 & \textbf{0.8785} \\\hdashline
 Op.66 & 660 & 197 & 0.7720 & 0.9439 & 0.8494 & 0.9425 & 0.9439 & \textbf{0.9432} \\\hdashline
 Op.69 No.2 & 591 & 186 & 0.7622 & 0.9272 & 0.8366 & 0.9649 & 0.7902 & \textbf{0.8688} \\\hdashline
 B.49 & 1111 & 441 & 0.7091 & 0.9172 & 0.7998 & 0.8175 & 0.9919 & \textbf{0.8963} \\
 \hline
 \textbf{Average} & 740 & 270 & 0.7392 & 0.9387 & 0.8262 & 0.8591 & 0.9396 & \textbf{0.8938} \\
 \hline
\end{tabular}
\end{table*}

\begin{figure*}[t]
\centering
\includegraphics[width=0.95\linewidth]{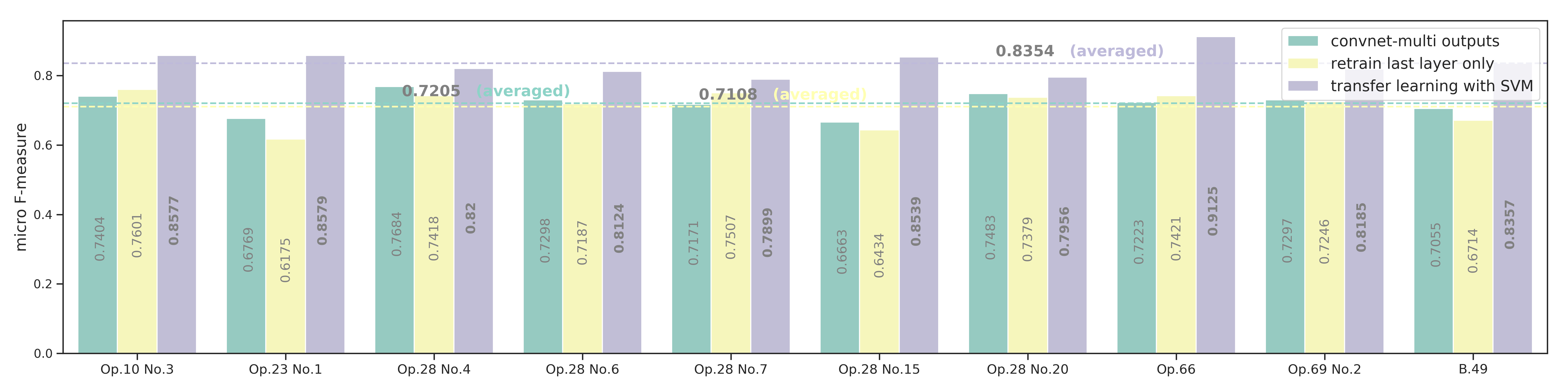}
\caption{Overall performance of the three methods in the target task.}
\label{fig:microf1}
\end{figure*}

In our experiments, melspectrograms with 128 mel bands were extracted from excerpts to serve as input to the network, The processing was done in real-time on the GPU using \textit{Kapre} \cite{choi2017kapre}, which can simplify audio preprocessing and saves storage. Time-frequency transformation was performed using 1024-point FFT with a hop size of 441 samples (10 ms). \textit{Keras} \cite{chollet2015keras} and \textit{Tensorflow} \cite{tensorflow2015-whitepaper} frameworks were used for the implementation.

\subsection{Source Task}
\label{sec:st}
The 62424 excerpts were split into 80\%/20\% to form the training/validation set. Models were trained until the validation accuracy no longer improved for 10 epochs. Batch size was set to 128 examples. To examine which \texttt{convnet} model can best discriminate \textit{pedal} versus \textit{no-pedal} excerpts, we compared best AUC-ROC scores (or simply AUC, representing Area Under Curve - Receiver Operating Characteristic) based on the validation set. 

As we introduced in Section \ref{sec:method}, we focused on the filter shape, i.e., ($m$, $n$) within the first layer. Models with the following ($m$, $n$) were trained:
\begin{itemize} 
\item As a baseline: (3, 3) (hereafter designated as \texttt{convnet-baseline}).
\vspace{5pt}
\item For modelling larger frequency contexts: (9, 3), (20, 3), (45, 3) (collectively denoted by \texttt{convnet-frequency}). These values of kernel length in frequency were motivated by the piano acoustics and physical structure, which fundamentally decide how the sustain-pedal effect sounds at notes of different registers. Since the mel scale was used, (9, 3) can at least cover 283 Hz, which approximately corresponds to the frequency of note \textit{C4}, a split point between bass and treble. Accordingly, (20, 3) and (45, 3) can be separately mapped to note \textit{D5} and \textit{G6}. The \textit{stress bar} near the strings of \textit{D5} separates the piano frame into different regions. The strings associated to notes higher than \textit{G6} are always free to vibrate because there are no more dampers above these strings.
\vspace{5pt}
\item Finally, for modelling larger time contexts: (3, 10), (3, 20), (3, 30), covering 100, 200 and 300 ms respectively (collectively denoted by \texttt{convnet-time}).
\end{itemize}

The number of channels ($c$) was set to 21 for all convolutional layers. Table \ref{tab:convnet} presents the accuracy and AUC scores of the above models obtained from the validation set. According to the best AUC score of \texttt{convnet-frequency} and \texttt{convnet-time} respectively, we selected (45,3) and (3,10) along with (3, 3) to create another model with multiple filter shapes (\texttt{convnet-multi}). To be specific, the first convolutional layer of \texttt{convnet-multi} consisted of (7, (45, 3)), (7, (3, 10)) and (7, (3, 3)). Its outputs were then concatenated along the channel dimension. The best accuracy and AUC scores were achieved by \texttt{convnet-multi}, i.e., 0.9837 and 0.9983. 

It is noted that all the models above obtained AUC score higher than 0.99 due to the relative simplicity of the classification task. To examine if the same level of performance can be obtained with fewer trainable parameters, we trained models similar to \texttt{convnet-multi} but with fewer channels and convolutional layers. According to the results in Table \ref{tab:convnetmulti}, the original \texttt{convnet-multi} remains the model with the highest score of AUC. Therefore, it was selected as the final model from the source task in order to be used as a feature extractor in the following target task.

\subsection{Target Task}
\label{sec:tt}
In the target task, sliding window was applied to the acoustic piano recordings in order to extract features of the trained \texttt{convnet-multi} model at every frame, as introduced in Section \ref{sec:tf}. The window covers a duration of 0.3 seconds with a hop size equivalent to 0.1 seconds. The 0.3-second samples were then tiled to 2 seconds and transformed into melspectrogram such that the input size was coherent with the one in the source task. The extracted features were used to train the SVM constructed by \textit{Scikit-learn} \cite{pedregosa2011scikit}.

The experiment was done by conducting {\em leave-one-group-out} cross-validation, where samples were grouped in terms of music passages. The performance of the proposed transfer learning method was validated in each music passage where the frame-wise features need to be classified by the SVM into pedal \textit{on} or \textit{off}, while the rest of the passages constitute the training set. The SVM parameters were optimised using grid-search based on the validation results. Radial kernel was used. Its bandwidth and the penalty parameter were selected from the ranges below:

\begin{itemize} 
\item bandwidth: [$1/2^3$, $1/2^5$, $1/2^7$, $1/2^9$, $1/2^{11}$, $1/2^{13}$, $1/\textit{feature vector dimension}$]
\item penalty parameter: [0.1, 2.0, 8.0, 32.0]
\end{itemize} 

We compared the proposed transfer learning method with the detection using a fine-tuned \texttt{convnet-multi} model, which can serve as a baseline classifier. Here ``fine-tuning'' is referred to as only retraining the fully-connected layer of \texttt{convnet-multi}. This is commonly considered a basic transfer learning technique. Within each cross-validation fold, the fully-connected layer was updated until the accuracy stopped increasing for 10 epochs. Then we obtained the fine-tuned \texttt{convnet-multi} outputs from short-time sliding windows over the melspectrogram of the validation passage.

Given the frame-wise \textit{on}/\textit{off} results for every music passage, we calculated precision ($P_1$), recall ($R_1$) and F-measure ($F_1$) with respect to the label \textit{on}. They are defined as:
\begin{equation*}
P_1 = \frac{N_{tp}}{N_{tp}+N_{fp}}, 
R_1 = \frac{N_{tp}}{N_{tp}+N_{fn}}, 
F_1 = 2 \times \frac{P_1 \times R_1}{P_1+R_1},
\end{equation*}
where $N_{tp}$, $N_{fp}$ and $N_{fn}$ are the numbers of true positives, false positives and false negatives respectively. 

Table \ref{tab:tt} presents the performance measurement of the two methods respectively for every validation passage in the cross-validation fold, where the occurrence counts of label \textit{on} and \textit{off} were obtained from the ground truth. In general, our proposed transfer learning method with SVM obtains better performance. We can observe that the average value of $P_1$ and $F_1$ are 11.99\% and 6.76\% higher in using the transfer learning method with SVM than with the fine-tuned \texttt{convnet-multi}. Both methods achieved similar average value of $R_1$.

We also compared the overall performance of the two methods along with directly using the pre-trained \texttt{convnet-multi}. Their results are presented  passage by passage in Figure \ref{fig:microf1}. Considering the imbalanced occurrence counts of the two labels, the micro-averaged F-measure ($F_{micro}$) was selected to evaluate the overall performance, because it calculates metrics globally by counting the total $N_{tp}$, $N_{fp}$ and $N_{fn}$ with respect to both labels. The proposed transfer learning method with SVM presents the best overall performance with more than 10\% higher than the $F_{micro}$ obtained by the other two methods.

% ==============================
%  Discussion
% ==============================
\subsection{Discussion}
\label{sec:discussion}
\begin{figure}[t]
\centering
\includegraphics[width=0.95\columnwidth]{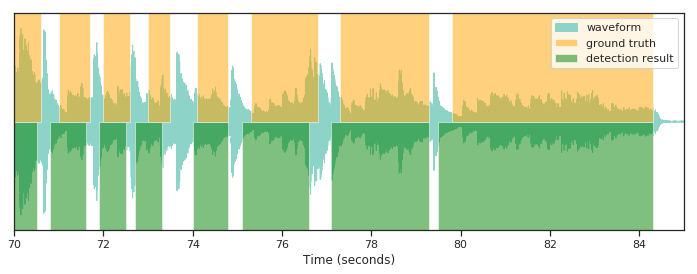}
\caption{Visualisation of the ground truth (top row) and the detection result (bottom row) in \textit{Op.66}. Audio frames that are annotated/detected as pedal \textit{on} are highlighted in orange/green.}
\label{fig:vis}
\end{figure}

\begin{figure*}[htp]
\centering
\subfloat[Melspectrograms of two input signals and their respective deconvolved results from 4 layers of \texttt{convnet-frequency}. \label{fig:f45t3}]{%
  \includegraphics[width=.95\linewidth]{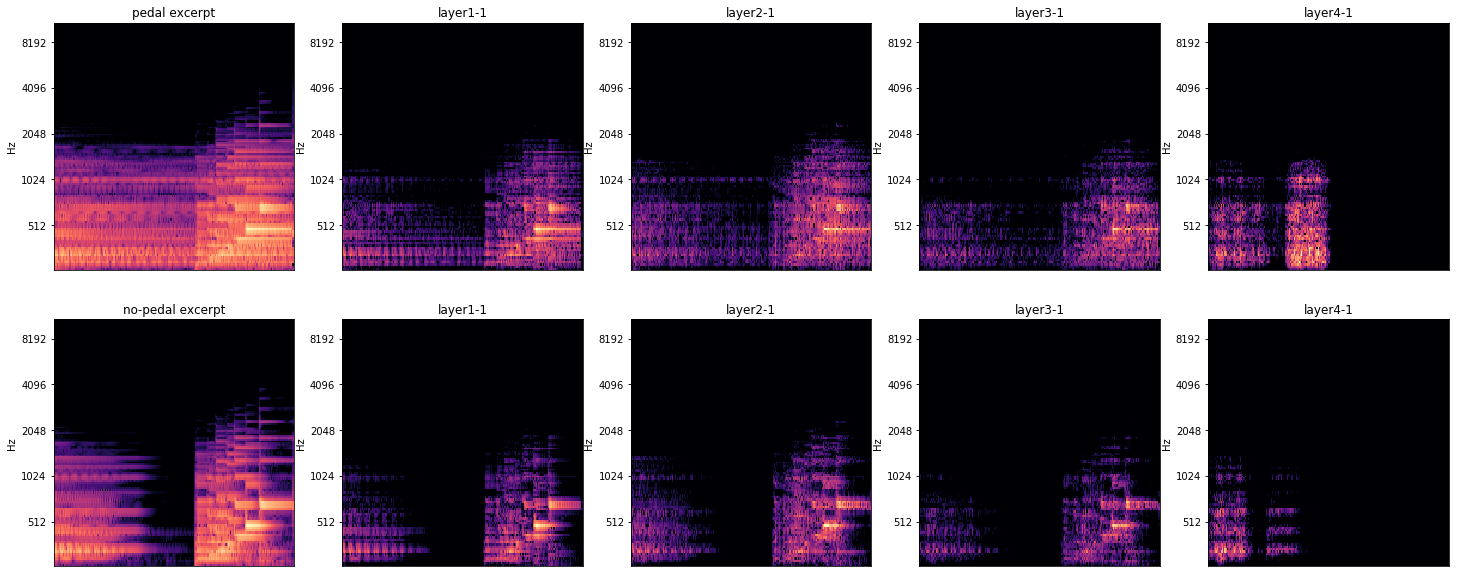}%
}

\subfloat[Melspectrograms of two input signals and their respective deconvolved results from 4 layers of \texttt{convnet-time}. \label{fig:f3t10}]{%
  \includegraphics[width=.95\linewidth]{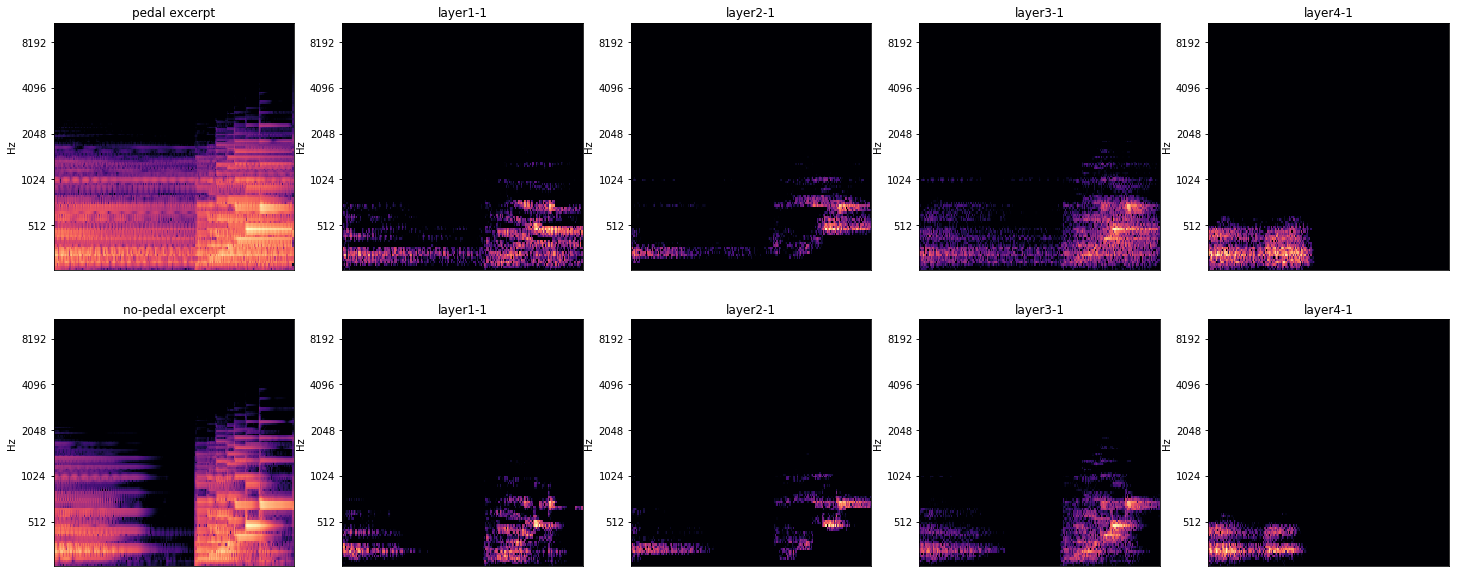}%
}
\caption{Visual analysis of music excerpts in pairs. The deconvolved melspectrogram corresponding to the first feature in layer $l$ is designated by \textit{layer$l$-1}.}
\label{fig:visft}
\end{figure*}

To gain deeper insight into the pros and cons of our method, we visualised the detection results of the last 15 seconds in the passage of \textit{Op.66}, which obtained the best performance in the target task. Figure \ref{fig:vis} highlights the audio frames corresponding to pedal \textit{on} according to the detection results and the ground truth separately. Most of the frames were correctly identified. Yet, there were false positives because some frames prior to the true sustain-pedal onset times were detected as positives, hence $P_1$ was decreased. This implies a model dedicated to the detection of the sustain-pedal onset should be developed. Such model itself or its outputs could be fused with \texttt{convnet-multi} in order to localise the pedalled segments with a better precision. There was also fragmentation corresponding to transient \textit{on} returned by the model, leading to increasing $N_{fp}$ and $N_{fn}$. This can be reduced by post-processing techniques.

It is notable in the source task that the AUC of models with various filter shapes within the first layer obtained scores that were all higher than 0.99 as shown in Table \ref{tab:convnet}. We assume that pressing the sustain pedal could result in acoustic characteristics that significantly change the patterns in both frequency and time. Thereby \texttt{convnet-multi} can obtain the highest AUC score. To understand the learning process of the \texttt{convnet} models, we conducted a visual analysis of the deconvolved melspectrogram of music excerpts in pairs, which have the same note event, but differently labelled. Visualisation results using the \texttt{convnet-frequency} and \texttt{convnet-time} with the best AUC score, i.e., with ($m$, $n$) set to (45, 3) and (3, 10), are shown in Figure \ref{fig:f45t3} and Figure \ref{fig:f3t10} respectively. In Figure \ref{fig:visft}, we only select the first feature maps separately learned in the four convolutional layers and present their deconvolved melspectrograms. From layer 1 to 3, the two models both focus on the time-frequency contexts centred around the fundamental frequency and their partials. More contexts in the higher frequency bands can be learned by the \texttt{convnet-frequency}. In the fourth layer, only the first half of melspectrograms are emphasised. We could infer the sustain pedal has more effects on the notes which the pedal just started to play together with. Meanwhile, the main differences between \textit{pedal} and \textit{no-pedal} excerpts lie in the lower frequency bands indicated by the \texttt{convnet-time}. Considering that a slightly lower accuracy score was obtained by \texttt{convnet-frequency}, we can assume dependencies within the higher frequency range could be a redundant knowledge to learn in our source task.

Another observation is that performance of the binary classification task is less dependent on the effect of the number of layers, according to the scores in Table \ref{tab:convnetmulti}. This also reflects in the changes shown by the deconvolved melspectrograms from layer 1 to 3, where roughly the same time-frequency areas were emphasised. We could train \texttt{convnet} models in a more efficient way, using fewer convolutional layers, while keeping or increasing the number of channels.

Through inspection of the detection results and the learned filters, we can extend our understanding of the \texttt{convnet} models in music. This inspires us to develop CNN models not only for detecting the pedalled frames, but also for learning the transients introduced by the sustain-pedal onset or even the offsets. More audio data including pieces by other composers and using various recording conditions should be tested to verify the robustness of our approach. This also constitutes our future works. 

% ==============================
%  Conclusion
% ==============================
\section{Conclusion}
\label{sec:conclusion}
In this paper, we answered the question: ``Can a computer point out pedalling techniques when a piano recording from a virtuoso performance is given?''. A novel transfer learning approach based on \texttt{convnet} models was proposed to detect the sustain pedal, and evaluated on ten passages of Chopin's music. A specific transfer learning paradigm was used where the source and target tasks differ in objectives and experimental conditions, including the use of synthesised versus real acoustic recordings. The model trained in the source task can then be employed as a feature extractor in the target task. 

In the source task, the model architecture was informed by piano acoustics and physics in order to facilitate the training process. Given the synthesised excerpts played with or without the sustain pedal, we showed that \texttt{convnet} models can learn the time-frequency contexts corresponding to acoustic characteristics of the sustain pedal, instead of larger variations introduced by other musical elements. Among all models, \texttt{convnet-multi} was selected to be used in the target task due to its highest scores of accuracy and AUC in binary classification. Features with more representation power dedicated to the sustain-pedal effect can be extracted from the intermediate layers of \texttt{convnet-multi}. This helps to adapt the detection to acoustic piano recordings. Thus a better performance measurement was obtained compared to fine-tuning or directly applying the pre-trained \texttt{convnet-multi} network. Finally, visualisation of the learned filters using deconvolution showed us potential directions towards designing more efficient and effective models for detecting different phases of the use of the sustain pedal.

% The \texttt{convnet} was trained to discriminate excerpts played with or without the sustain pedal using melspectrograms in pairs from a synthesised dataset. This enabled \texttt{convnet} to learn the time-frequency contexts corresponding to acoustic characteristics of the sustain pedal instead of pitch variance. We experimented with different filter shapes within the first layer in order to train a \texttt{convnet} with the best discriminating ability. The \texttt{convnet-multi} was thereby selected. Its activations from every convolutional layers were applied an average-pooling and then concatenated to form the features to be transferred. Such features from every frame of the music passages can be classified by SVM into pedal \textit{on} or \textit{off}. We adopted {\em leave-one-group-out} cross-validation to compare the proposed method with directly using the output of the pre-trained \texttt{convnet-multi}. A better performance measurement was obtained by the transfer learning approach. Moreover, visualisation of the learned filters using deconvolution showed us potential focuses on designing more efficient models related to sustain-pedal detection in the future.

% \balance
\bibliographystyle{IEEEtran}
\bibliography{IEEEabrv,ref}
\end{document}